\documentclass[sigconf]{acmart}

\AtBeginDocument{%
  }

\setcopyright{acmlicensed}
\copyrightyear{2018}
\acmYear{2018}
\acmDOI{XXXXXXX.XXXXXXX}

\acmConference[FSE '25]{Companion Proceedings of the 33rd ACM Symposium on the Foundations of Software Engineering}{June 23--27,
  2025}{Trondheim, Norway}
\acmBooktitle{Companion Proceedings of the 33rd ACM Symposium on the Foundations of Software Engineering (FSE '25), June 23--27, 2025, Trondheim, Norway}
\acmISBN{978-1-4503-XXXX-X/2018/06}

\makeatletter
\newcommand{\removelatexerror}{\let\@latex@error\@gobble}
\makeatother

\usepackage{microtype}
\usepackage{url}                  
\usepackage{listings}          

\makeatletter
\newcommand\BeraMonottfamily{%
  \def\fvm@Scale{0.75}
  \fontfamily{fvm}\selectfont
}
\makeatother

\usepackage{enumitem}      

\usepackage[htt]{hyphenat}
\usepackage{graphicx}
\usepackage{float,subfig}
\usepackage{xspace,framed}
\usepackage{colortbl}
\usepackage{calc}
\usepackage[ruled,vlined]{algorithm2e}

\usepackage{amsthm}
\newtheorem{definition}{Definition}
\newtheorem{example}{Example}

\usepackage{amsfonts}
\usepackage{hyperref}
\usepackage{multicol}
\usepackage{multirow}
\usepackage{tikz}
\usepackage[justification=centering]{caption}
\usepackage{stmaryrd}
\usetikzlibrary{positioning, arrows, automata, shapes}
\usepackage{hhline}
\usepackage{pifont}
\usepackage{pdflscape} 
\usepackage{longtable}
\usepackage{afterpage}
\usepackage{wasysym}
\usepackage{tcolorbox}
\usepackage{float}
\definecolor{prompt_bg}{RGB}{252,255,221}
\definecolor{prompt_title}{RGB}{0,51,102}
\newcommand{\templatevar}[1]{\$\{\textcolor{red}{#1}\}}

\usepackage[justification=centering]{caption}

\newcommand{\doi}[1]{}{}

\newcommand{\tool}{\textsc{RefSym}\xspace}
\newcommand{\llm}{\textsc{RefNeural}\xspace}
\newcommand{\llma}{\textsc{RefNeural}-Claude2.1\xspace}
\newcommand{\llmb}{\textsc{RefNeural}-Claude3\xspace}

\newcommand{\var}[1]{\ensuremath{\mathit{{#1}}}}

\newcommand*{\codenospace}[1]{\lstinline|#1|\kern-1ex}%

\newcommand{\xmark}{\ding{55}}%

\tikzset{
    box/.style={
           rectangle,
           rounded corners,
           draw=black, very thick,
           fill=yellow!20,
           text width=10em,
           minimum height=2em,
           text centered},
  futurebox/.style={
           rectangle,
           rounded corners,
           draw=black, very thick,
           fill=blue!10,
           text width=10em,
           minimum height=2em,
           text centered},
  longfuturebox/.style={
           rectangle,
           rounded corners,
           draw=black, very thick,
           fill=purple!20,
           text width=10em,
           minimum height=2em,
           text centered},
    arrow/.style={
           ->,
           thick,
           shorten <=2pt,
           shorten >=2pt,}
}

\allowdisplaybreaks

\newcommand{\consumableobjs}{\mbox{$\mathit{consumable\_objs}$}\xspace}
\newcommand{\availabletypes}{\mbox{$\mathit{available\_types}$}\xspace}
\newcommand{\targettypes}{\mbox{$\mathit{target\_types}$}\xspace}
\newcommand{\requiredtypes}{\mbox{$\mathit{required\_types}$}\xspace}

\begin{document}





\title{Quantifying the benefits of code hints\\ for refactoring deprecated Java APIs }



\author{Cristina David}
\affiliation{%
  \institution{University of Bristol}
  \city{Bristol}
  \country{UK}}

\author{Pascal Kesseli}
\affiliation{%
  \institution{Meta}
  \city{Zurich}
  \country{Switzerland}
}

\author{Daniel Kroening}
\affiliation{%
  \institution{Amazon Inc.}
  \city{Seattle}
 \country{US}}

\author{Hanliang Zhang}
\affiliation{%
  \institution{University of Bristol}
  \city{Bristol}
  \country{UK}}

\copyrightyear{2025}
\acmYear{2025}
\setcopyright{cc}
\setcctype{by}
\acmConference[FSE Companion '25]{33rd ACM International Conference on the
Foundations of Software Engineering}{June 23--28, 2025}{Trondheim, Norway}
\acmBooktitle{33rd ACM International Conference on the Foundations of
Software Engineering (FSE Companion '25), June 23--28, 2025, Trondheim,
Norway}\acmDOI{10.1145/3696630.3728567}
\acmISBN{979-8-4007-1276-0/2025/06}




\begin{abstract}
When done manually by engineers at Amazon and other companies, refactoring legacy code in order to eliminate uses of
deprecated APIs is an error-prone and time-consuming process.  In this
paper, we investigate to which degree refactorings for deprecated Java APIs
can be automated, and quantify the benefit of Javadoc code hints for this
task.  To this end, we build a symbolic and a neural engine for the
automatic refactoring of deprecated APIs.  The former is based on
type-directed and component-based program synthesis, whereas the latter uses
LLMs.  We applied our engines to refactor the deprecated methods in the
Oracle JDK 15.  Our experiments show that code hints are enabling for the
automation of this task: even the worst engine correctly refactors 71\% of
the tasks with code hints, which drops to at best 14\% on tasks without.
Adding more code hints to Javadoc can hence boost the refactoring of code
that uses deprecated APIs.
\end{abstract}

\ccsdesc[500]{Software and its engineering~Empirical software validation}
\ccsdesc[500]{Computing methodologies~Machine learning}
\ccsdesc[500]{Software and its engineering~Source code generation}

\keywords{Program Refactoring, Program Synthesis, LLMs}

\maketitle





\section{Introduction}\label{sec:intro}






As programming languages evolve, industry engineers take on the critical task of migrating code to newer versions, an important aspect of code modernisation. This challenge spans a wide range of companies, including Amazon. Such migrations ensure access to new features, improved performance, and enhanced security while maintaining compatibility with modern tools and libraries. Java, for example, has frequent JDK releases, each introducing improvements and updates. However, migrating to newer JDK versions presents challenges, particularly due to deprecated APIs. Deprecation, intended to enhance security, performance, and maintainability, requires engineers to update existing code by replacing deprecated APIs with modern, more efficient alternatives, adding complexity to the migration process. 





The transformation of existing code such that it doesn't use
deprecated APIs is not always straightforward, as illustrated in
Figure~\ref{ex:deprecated-method-other}. In the example, we make use of
the \lstinline[breaklines=true]{getHours} method of the \lstinline[breaklines=true]{Date} class, which is
deprecated.  In this situation, in order to replace the use of the
deprecated method, we must first obtain a \lstinline[breaklines=true]{Calendar} object.
However, we can't use the \lstinline[breaklines=true]{Calendar} constructor as it is
protected, and we must instead call \lstinline[breaklines=true]{getInstance}.
Furthermore, in order to be able to use this \lstinline[breaklines=true]{Calendar} object
for our purpose, we must first set its time using the existing
\lstinline[breaklines=true]{date}.  We do this by calling \lstinline[breaklines=true]{setTime} with
\lstinline[breaklines=true]{date} as argument.  Finally, we can retrieve the hour by
calling \lstinline[breaklines=true]{calendar.get(Calendar.HOUR_OF_DAY)}.

\begin{figure}[ht]
\begin{lstlisting}[mathescape=true,showstringspaces=false,basicstyle=\small]
void main(String[] args) {
  // Deprecated:
  int hour=date.getHours();

  // Should have been:
  final Calendar calendar=Calendar.getInstance();
  calendar.setTime(date);
  int hour=calendar.get(Calendar.HOUR_OF_DAY);
}
\end{lstlisting}
\caption{Deprecated method example.}
\label{ex:deprecated-method-other}
\end{figure}

%



Refactoring code that relies on deprecated APIs presents additional challenges. For instance, the code to be refactored might be using abstract classes and abstract methods, and it may not be obvious how to subclass from the code to be refactored (e.g.~\lstinline[breaklines=true]{engineGetParameter} in \lstinline[breaklines=true]{java.security.SignatureSpi}). It may also call methods that, while not abstract, must be overridden by subclasses (e.g., the \lstinline[breaklines=true]{layout} method in \lstinline[breaklines=true]{java.awt.Component}, which has an empty body). Furthermore, the code might invoke native methods implemented in other programming languages, such as C or C++ (e.g., \lstinline[breaklines=true]{weakCompareAndSet} in \lstinline[breaklines=true]{java.util.concurrent.atomic.AtomicReference}). Understanding the behavior of such code requires knowledge of how to subclass abstract classes, override methods appropriately, and understand the functionality of native code written in other languages.


Manual refactoring of deprecated APIs is a time-consuming and error-prone process. \emph{Therefore, in this paper, we investigate the automatic generation of refactorings for deprecated APIs.} While our primary emphasis is on refactoring deprecated methods, the same techniques can be extended to handle deprecated fields and classes.

When deprecating a field, method, or class, the
\lstinline[breaklines=true]{@Deprecated} Javadoc tag is used in the comment
section to inform the developer of the reason for deprecation and,
sometimes, what can be used in its place.  We call such a suggestion a {\em
  code hint}. These hints are helpful for a human performing the refactoring.
While they don’t offer a complete solution, they serve as valuable guidance, making the process more intuitive

For illustration, let's look at our running example in
Figure~\ref{ex:deprecated-method-other}.  The source code for the
\lstinline[breaklines=true]{getHours} method in class \lstinline[breaklines=true]{Date} is accompanied by the
comment in Figure~\ref{ex:code-hints}, where the \lstinline[breaklines=true]{@code} tag suggests
replacing the deprecated \lstinline[breaklines=true]{getHours} by
\lstinline[breaklines=true]{Calendar.get(Calendar.HOUR_OF_DAY)}.
%
Using this code hint is not straightforward.
Although it may seem as if
method \lstinline[breaklines=true]{get} is static, allowing an immediate call,
it is actually an instance method, and requires an object of
class \lstinline[breaklines=true]{Calendar}. However, no such object is available in the
original code, meaning that it must be created by the refactored code.
Consequently, we must find the necessary
instructions that consume existing objects and create a \lstinline[breaklines=true]{Calendar} object.
Besides generating the required objects, we also need to set their
fields. For instance,
in Figure~\ref{ex:deprecated-method-other}, we must call
\lstinline[breaklines=true]{calendar.setTime(date)} to set the calendar's date
based on the existing \lstinline[breaklines=true]{date} object.

Given that code hints are useful to humans,
\emph{in this paper, we are also interested in investigating the benefits of Javadoc code hints in the automation of the refactoring.}

\begin{figure}
\begin{lstlisting}[mathescape=true,showstringspaces=false]
/*
 * @deprecated As of JDK version 1.1,
 * replaced by {@code Calendar.get
 *              (Calendar.HOUR_OF_DAY)}.
 */
\end{lstlisting}
\caption{Code hints for the running example.}
\label{ex:code-hints}
\end{figure}




In order to investigate the benefits of code hints when automating the deprecated refactoring,
we build two code generation engines, namely a symbolic and a neural one.  For each, we selected approaches with proven success in program synthesis, and that could also incorporate code hints.
%
In particular, for the symbolic engine, we make use of component-based synthesis~\cite{DBLP:conf/icse/JhaGST10,DBLP:conf/pldi/GulwaniJTV11,DBLP:conf/popl/FengM0DR17}
and type-directed synthesis~\cite{DBLP:conf/sfp/Katayama05,DBLP:conf/pldi/OseraZ15,DBLP:journals/pacmpl/YamaguchiMDW21}. Essentially, we use types and code hints to populate
a component library (i.e., a library of instructions), such that these components are then weaved together to generate the desired program.
Given the recent success of Large Language Models (LLMs) for code generation~\cite{modular,DBLP:journals/corr/abs-2302-05527,DBLP:journals/corr/abs-2202-13169,llmsforcodecompletion,codegenclasslevel,zhang2023repocoder,lostintranslation,yang2024sweagent}, the neural approach
generates candidate refactorings by iteratively querying an LLM -- we used Claude 2.1 and Claude 3~\cite{claude}.


Our experiments show that code hints are important for the performance of
both the symbolic and the neural engines: when code hints are given, both
engines perform very well, making full automation of deprecated APIs
refactorings feasible.  Without code hints, the refactoring becomes much
harder for both engines, such that the vast majority of benchmarks without
code hints are failing.  Thus, our conclusion is that, in order to
facilitate the automation of the refactoring of client code that uses
deprecated APIs, code hints should be added to all deprecated methods that
have a replacement.

Comparing the symbolic and the neural approaches, code hints help the
symbolic approach to efficiently prune the solution space, resulting in a
better performance than the neural engine at a lower computational cost.  We
believe this is important to note, especially as LLMs are often seen as a
panacea for all tasks, including code generation.  This work shows that
symbolic methods can still be effective in specialised settings, where
efficient pruning of the solution space is possible.

\paragraph{Contributions}

\begin{itemize}

\item We propose a symbolic and a neural refactoring technique for deprecated APIs. The former makes use of type-directed and component-based synthesis, whereas the latter is LLM based. In order to check the correctness of the refactorings, we design a symbolic equivalence check, which takes into consideration both the state of the stack and the heap.


\item We implement our techniques and use them to refactor deprecated methods from the Oracle
JDK 15 Deprecated API documentation~\cite{OracleJdk15DeprecatedAPI}.

\item We investigate the benefits of code hints for both the symbolic and neural approach. 
  Our results show that code hints are important for the performance of both the symbolic and the neural engines: adding more code hints to Javadoc
can substantially help automate the refactoring of deprecated APIs.

\end{itemize}

\section{Our approach} \label{sec:overview}
In this section, we describe the two code generation engines.
For both approaches, we make use of a CounterExample Guided Inductive Synthesis (CEGIS)~\cite{DBLP:conf/pldi/Solar-LezamaJB08} architecture, where we
iteratively attempt to improve a candidate refactoring
until it behaves indistinguishably
from the original code, i.e., the original and
the refactored blocks of code are observationally equivalent.
In the rest of the paper, we will use the predicate
$equivalent(P_1(\vec{i}), P_2(\vec{i}))$ to denote that code $P_1$ is
observationally equivalent to $P_2$ for inputs $\vec{i}$, meaning that the two programs
can't be distinguished by their behaviour on inputs $\vec{i}$.  In general,
we refer to the original code as $P_1$ and the refactored code as
$P_2$. We will fully define what $equivalent$ actually means in Section~\ref{sec:equiv}.

In each iteration of the synthesis process, there are two phases, a synthesis phase and a verification phase. The synthesis phase generates a {\em candidate refactoring} that is equivalent to
the original code on a finite set of inputs. Then, the verification phase tries to find a new {\em counterexample input} that distinguishes between the current candidate refactoring and the original code. If it manages, this input is added to the set of finite inputs used by the next synthesis phase, and if it fails, then the current candidate is indeed a solution refactoring.


\subsection{Synthesis phase}\label{sec:synthesis}
In this phase, we have
a finite set of input examples $\{\vec{i_1} \cdots \vec{i_n}\}$ and we attempt to find a candidate refactoring
that is observationally equivalent to the original code for the given inputs.

\paragraph{Symbolic engine}
We construct the following \lstinline[breaklines=true]{Synthesise} method, which
takes the refactored code $P_2$ as input.

\begin{lstlisting}[mathescape=true,showstringspaces=false,numbers=none]
Synthesise ($P_2$) {
  if (equivalent($P_1$($\vec{i_1}$), $P_2$($\vec{i_1}$)) && ...
      && equivalent($P_1$($\vec{i_n}$), $P_2$($\vec{i_n}$)))
    assert(false);
}
\end{lstlisting}

This method consists of only one conditional saying that if $P_1$ and $P_2$ are equivalent on all given inputs $\vec{i_1}, \cdots, \vec{i_n}$, then \lstinline[breaklines=true]{assert(false)} is reached.
Given that this assertion always fails if reached, it means that the method is unsafe when $P_1$ and $P_2$ are equivalent on the given inputs.
Thus, we can reduce the synthesis problem to the problem of checking the safety of \lstinline[breaklines=true]{Synthesise}. If a safety checker manages to
find an input $P_2$ for which the assertion fails (meaning that \lstinline[breaklines=true]{Synthesise} is unsafe), then this $P_2$ must be equivalent to $P_1$ on the given inputs. This $P_2$ is the {\em refactoring candidate}
that the synthesis phase is supposed to generate.

We use an existing fuzz testing platform for Java, JQF~\cite{DBLP:conf/issta/PadhyeLS19},
as the safety checker. JQF is designed to handle structured inputs, where inputs of type \lstinline[breaklines=true]{T}
are generated by a backing \lstinline[breaklines=true]{Generator<T>}. JQF provides a library of
generators for basic types such as primitive values. We implement custom
JQF generators based on a library of instructions built as explained in Section~\ref{sec:components-seeding},
enabling JQF to construct $P_2$ by weaving together instructions from the library.

If the safety checker fails to find a $P_2$ for which the assertion fails, then the overall synthesis technique fails to generate a solution refactoring.
There could be two reasons for this situation, which we cannot differentiate between.
Firstly, there may indeed be no $P_2$ that can be constructed with the available components in the component library such that
it is equivalent to $P_1$ on the given inputs. Secondly, this could be caused by JQF's unsoundness.
Given that fuzzers rely on testing, they may fail to find inputs that trigger unsafe behaviours even when such inputs exist.
Consequently, we cannot guarantee that we will always find a refactoring whenever one exists. We will provide more details about our decision to use fuzzing as the safety checker in
Section~\ref{sec:implementation}.


\paragraph{Neural engine}
For each deprecated method, we query the LLM by constructing a
prompt as given in \autoref{fig:prompt}.  The method definition and the
JavaDoc comment (if present) are provided as context.  Furthermore, we
provide a code snippet of the use of the deprecated method that is to be
refactored (e.g., \lstinline{this.minimumSize();}).  Finally, we attach a
list of formatting constraints.

The code returned by the LLM is verified against the original code.
If the verification fails, we attach the set of counterexamples generated by the fuzzing
engine to subsequent prompts. We apply object serialisation to obtain structured-views of the input and output states.
In addition to this, we followed the prompting guidelines provided by Anthropic Claude\footnote{https://docs.anthropic.com/en/docs/prompt-engineering} (e.g. using XML tags).
Do note that, as opposed to the symbolic approach, the LLM doesn't give any guarantees that the counterexamples were actually taken into consideration.

\begin{figure}
  \centering
  \begin{tcolorbox}[
    colback=prompt_bg,
    colframe=prompt_title,
    subtitle style={boxrule=0.4pt, colback=yellow!50!blue!25!white, colupper=black}
  ]
  \scriptsize

  \tcbsubtitle{Initial Prompt}
  \textbf{User:}\\
  \textit{\# Context}\\
  \texttt{The method \templatevar{METHOD\_NAME} of the class \templatevar{CLASS\_NAME} is deprecated. Below is the method definition:}\\
  \texttt{\templatevar{METHOD\_DEFINITION}}\\

  \texttt{Here are its Javadoc comments that may contain a @deprecated tag explaining why the item has been deprecated and suggesting what to use instead.}\\
  \texttt{\templatevar{JAVADOC\_COMMENT}}\\

  \texttt{However, I used this method call in my code base, the code snippet is given below:}\\
  \texttt{\templatevar{CODE\_SNIPPET}}\\

  \textit{\# Instruction}\\
  \texttt{Help me refactor this code snippet so that it doesn't use the deprecated method. Do not simply inline the method body, use APIs suggested by the Javadoc comments if there are any.}\\

  \textit{\# Constraints}\\
  \texttt{Take the following constraints into consideration:}\\
  \texttt{\templatevar{FORMATTING\_CONSTRAINTS}}\\

  \tcbsubtitle{Subsequent Prompt}
  \texttt{\textcolor{gray}{Additionally, here is a set of input/output examples that you should respect,}}\\
  \texttt{\templatevar{EXAMPLES}}\\

  \textbf{Assistant:}\\
  \end{tcolorbox}
  \caption{LLM Prompt Template.}
  \label{fig:prompt}
\end{figure}

\subsection{Verification phase}
This is exactly the same for the two approaches.
We are provided with a candidate
refactoring $P_2$ and we must check whether there exists any input
$\vec{i}$ for which the original code and the candidate
refactoring are not observationally equivalent.  To do this, we build
the following \lstinline[breaklines=true]{Verify} method, which given some input
$\vec{i}$, asserts that the two programs are equivalent for
$\vec{i}$.

\begin{lstlisting}[mathescape=true,showstringspaces=false,numbers=none]
Verify($\vec{i}$) { assert(equivalent($P_1(\vec{i})$, $P_2(\vec{i})$)); }
\end{lstlisting}

In other words, answering the question posed by the verification
phase is reduced to checking the safety of this method: if
\lstinline{Verify} is safe (i.e., the assertion is not violated) for any
input~$\vec{i}$, then there is no input that can distinguish
between the original code and the candidate refactoring (this
is indeed a sound refactoring). However, if \lstinline[breaklines=true]{Verify} is not safe,
then we want to be able to obtain a {\em counter\-example input}
$\vec{i_{cex}}$ for which the assertion fails.
This counterexample will be provided back to the synthesis phase and
used to refine the current candidate refactoring. Again, we check the
safety of \lstinline[breaklines=true]{Verify} with JQF.

In this section, we hid the complexity of the equivalence check inside
the $equivalent$ predicate. This is far from trivial as it requires handling
of notions such as aliasing, loaded classes, static fields etc. This will be
described in detail next.

\subsection{Checking program equivalence}\label{sec:equiv}

A core part of the synthesis procedure is the $equivalent(P_1(\vec{i}), P_2(\vec{i}))$ predicate,
which checks that the original code $P_1$ and a candidate refactoring $P_2$
are equivalent for a given input $\vec{i}$. 
In this section, we provide details on how we check this equivalence.

The state of a Java program is modelled by the current program stack
(consisting of method-specific values and references to objects in the
heap) as well as its heap (consisting of instance variables and static
field values). Static field values are stored with their respective
classes, which in turn are loaded by class loaders.  Our equivalence
check must take all these into consideration. One of the main challenges is aliasing, which we will discuss later in the section.

We start by introducing some notation:
\begin{itemize}
\item \var{loadedClasses(P)} returns the set of classes loaded by the class loader
  in which \var{P} is executed. Note that Java allows to load the same class
in different class loaders, which creates independent copies of its
static fields.
\item \var{liveVars(P)} provides the set of variables that are live
at the end of $P$. 
\item \var{staticFields(C)} returns
  the set of static fields of class \var{C}.
\item $exc(P, \vec{i})$ returns the exception thrown by $P$'s execution on input $\vec{i}$.
\end{itemize}

\begin{example}\label{ex:defs}
  For our running example in Figure~\ref{ex:deprecated-method-other},
  \var{P_1} and \var{P_2} are represented by the following lines of code. Note that
  both $P_1$ and $P_2$ take variable $date$ as input.

\begin{lstlisting}[mathescape=true,showstringspaces=false]
  // $P_1$
  int hour=date.getHours();

  // $P_2$
  final Calendar calendar=Calendar.getInstance();
  calendar.setTime(date);
  int hour=calendar.get(Calendar.HOUR_OF_DAY);
\end{lstlisting}
Then, we have:
\[
\small
\begin{aligned}[t]
  \var{liveVars(P_1)} &= \{date, hour\}\\
  \var{liveVars(P_2)} &= \{calendar, date, hour\}\\
  \var{loadedClasses(P_1)} &= \{Date\} \\
  \var{loadedClasses(P_2)} &= \{Calendar, Date\} \\
  \var{staticFields(Date)} &= \emptyset\\
  \var{staticFields(Calendar)} &= \{DATE, YEAR, \cdots\}\\
  \var{exc(P_1, \_)}=\var{exc(P_2, \_)} &=\emptyset
\end{aligned}
\]
\end{example}


In order to extract the (last) object assigned to a variable $v$ by the execution of $P$ on a specific input $\vec{i}$,
we will use the notation $E[P(\vec{i})](v)$. Essentially, if we consider the trace generated by executing $P(\vec{i})$,
then $E[P(\vec{i})]$ maps each variable defined in $P$ to the last object assigned to it by this trace.
Next, we define the notion of equivalence with respect to a concrete input $\vec{i}$ (this definition is incomplete and we will build on it in the rest of the section).
We overload equality to work over sets (for live variables and loaded classes).

\begin{definition}[Program equivalence with respect to a concrete input $\vec{i}$ {\bf[partial]}]\label{def:prog-equiv}
  Given two code blocks $P_1$ and $P_2$ and concrete input~$\vec{i}$,
  we say that \var{P_1} and \var{P_2} are equivalent
  with respect to $\vec{i}$, written as $equivalent(P_1(\vec{i}), P_2(\vec{i}))$
  if and only if the following conditions hold:
%
%
  \[
\small
\begin{aligned}
  & (1)~ \var{exc}(P_1, \vec{i})=\{e_1\} \wedge \var{exc}(P_2, \vec{i})=\{e_2\} \wedge \var{equals}(e_1,e_2) ~\vee\\
  & \qquad \var{exc}(P_1, \vec{i})=\var{exc}(P_2, \vec{i}) =\emptyset \\
      & (2)~ \var{liveVar}(P_1) = \var{liveVar}(P_2) \wedge \forall v\in \var{liveVar}(P_1). \\
      & \qquad \var{equals}(E[P_1(\vec{i})](v), E[P_2(\vec{i})](v))\\
  & (3)~  \var{loadedClasses}(P_1) = \var{loadedClasses}(P_2) ~\wedge\\
  & \qquad \forall C \in \var{loadedClasses}(P_1). \forall f {\in} \var{staticFields}(C).\\
  & \qquad \var{equals}(E[P_1(\vec{i})](f), E[P_2(\vec{i})](f))
    \end{aligned}
    \]

  \end{definition}


The above definition says that in order for \var{P_1} and
\var{P_2} to be equivalent with respect to input $\vec{i}$,
(1) either they both throw an exception and the two exceptions have the same type, or none does,
(2) the set of variables live at the end of \var{P_1} is the same as the set of variables live at the end of  \var{P_2}, and must be
assigned equal objects by the executions of \var{P_1} and
\var{P_2} on $\vec{i}$, respectively, and
(3) the classes loaded by \var{P_1} must be the same as the classes loaded by \var{P_2}, and
all the static fields in the classes loaded by both the class loaders of \var{P_1} and \var{P_2} must be assigned equal objects by the
two executions, respectively.
In our implementation, if either the deprecated or the refactored code didn't load a class, we load it for it,
and check that the initial state of that class is the same for both blocks of code.

Equality refers to value equality, which we
check by recursively following attribute chains until we reach
primitive types.
We generally do not consider existing \lstinline[breaklines=true]{equals} methods
unless (i)~the method was not written by the user (i.e.,  JCL classes),
(ii)~the type inherits from \lstinline[breaklines=true]{java.lang.Object}, (iii)~the class
implements \lstinline[breaklines=true]{java.lang.Comparable} and (iv)~the type declares an
\lstinline[breaklines=true]{equals} implementation.  Examples of classes that satisfy these
strict requirements are \lstinline[breaklines=true]{java.lang.Integer} or
\lstinline[breaklines=true]{java.util.Date}.  All other implementations of \lstinline[breaklines=true]{equals} are
considered unreliable and ignored.




\begin{example}\label{ex:equiv}
  When applying Definition~\ref{def:prog-equiv} to \var{P_1} and \var{P_2} given in Example~\ref{ex:defs},
  the following conditions must hold (given that class \var{Date} has no static fields, the third condition is trivial):
  \[
\small
\begin{aligned}
      & (1)~ \var{exc(P_1)}=\var{exc(P_2)} =\emptyset\\
      & (2)~ equals(E[P_1(\vec{i})](hour), E[P_2(\vec{i})](hour)) ~ \wedge\\ 
      & \qquad equals(E[P_1(\vec{i})](date), E[P_2(\vec{i})](date))\\
& (3)~  \var{Date} \in \{\var{Date}, \var{Calendar}\}
    \end{aligned}
    \]


\end{example}


{\bf The challenge of aliasing.}
When expressing the equivalence relation between \var{P_1} and \var{P_2}, we
intentionally missed one important aspect, namely aliasing.  To understand
the problem let's look a the following example, which makes use of
java.awt.Container, where a generic Abstract Window Toolkit (AWT) container
object is a component that can contain other AWT components.  Method
\lstinline[breaklines=true]{preferredSize} used by the original code below returns the preferred
size of the calling container. Method
\lstinline[breaklines=true]{preferredSize} is deprecated, and, instead, the refactored version
uses \lstinline[breaklines=true]{getPreferredSize}.


\begin{example}\label{ex:aliasing}
~\begin{lstlisting}[mathescape=true,showstringspaces=false,numbers=none]
  // $P_3$:
  Dimension dim1=container.preferredSize();
  Dimension dim2=container.preferredSize();

  // $P_4$:
  Dimension dim1=container.getpreferredSize();
  Dimension dim2=dim1;
\end{lstlisting}







We note that, the original code above defines two variables \var{dim1} and
\var{dim2}, each assigned an object of type \lstinline[breaklines=true]{Dimension} returned by calling
\var{container.preferredSize()}.
Conversely, in the refactored code, \var{dim1} and \var{dim2} are aliases,
i.e., they point to the same
object of type \lstinline[breaklines=true]{Dimension}.

The original and the refactored code are equivalent according to Definition~\ref{def:prog-equiv}.
Let's next assume that the following code
is used after both the original and the refactored code, respectively.

\begin{lstlisting}[mathescape=true,showstringspaces=false]
  dim1.setSize(1,2);
  dim2.setSize(2,3);
\end{lstlisting}

If the original and the refactored code were indeed equivalent,
then we would expect \var{dim1} and \var{dim2} to have the same value at the end of both blocks of code.
However, this is not the case.
In the original code, \var{dim1} will have width 1 and height 2, whereas in the refactored code,
\var{dim1} will have width 2 and height 3. This is due to the fact that,
in the refactored code, \var{dim1} and \var{dim2} are aliases. 
Thus, when \var{dim2} has its size set to (2,3), this also affects \lstinline[breaklines=true]{dim1}.
\end{example}

Intuitively, any aliases between live variables at the end of the original code should also
be present at the end of the refactored code. For this purpose, we use the notation
\var{aliasEquivClass(V)} which returns the set of all equivalence classes induced over the
set of variables \var{V} by the aliasing relation.
Notably, the aliasing equivalence relation must also hold over the
static fields of the classes loaded by both programs.

\begin{example}
For \var{P_1} and \var{P_2}, there are no aliases.
%
However, for \var{P_3} and \var{P_4} we have:
\[
\small
\begin{aligned}[t]
  &\var{aliasEquivClass(liveVar(P_3)}~ \cup\\
  &\var{staticFields(loadedClasses(P_3))} = \emptyset\\
  &\var{aliasEquivClass(liveVar(P_4)} ~\cup\\
  &\var{staticFields(loadedClasses(P_4))} = \{\{dim1, dim2\}\}
\end{aligned}
\]
In \var{P_4}, the aliasing relation induces one equivalence class, namely \var{\{dim1, dim2\}}.

\end{example}

Next, we complete Definition~\ref{def:prog-equiv} by capturing the aliasing aspect.

\begin{definition}[Addition to Definition~\ref{def:prog-equiv}]\label{def:prog-equiv-add}
In addition to Definition~\ref{def:prog-equiv},
two programs \var{P_1} and \var{P_2} are equivalent
with respect to $\vec{i}$ iff:
\[
\small
    \begin{aligned}
      & (4)~ \forall v_1,v_2 \in \mathit{liveVar}(P_1) \cup \mathit{staticFields}(loadedClasses(P_1)). \\
      & \qquad aliases(P_1, v_1, v_2) \Longleftrightarrow  aliases(P_2, v_1,v_2)
    \end{aligned}
    \]
where, given a program $P$, variables $v_1$ and $v_2$ are aliases, i.e., $aliases(P, v_1, v_2)$ holds
if and only if they are in the same equivalence class induced by the aliasing relation,
  i.e., $aliasEquivClass(\allowbreak \{v_1,v_2\}) = \{\{v_1,v_2\}\}$.
\end{definition}

  We abuse the notation to use \var{staticFields} over a set of classes, rather than just one class.
  The objective is to return the union of all static fields defined in all the classes in the set of classes taken as argument.




\section{Seeding of the component library} \label{sec:components-seeding}

In this section, we describe how we populate the instruction library (also referred to as the component library)
used by the symbolic approach starting from code hints. We'll refer to this library as the CodeHints-library.
This step is critical as, in order for the symbolic engine to succeed, the library must be as small as possible
while containing all the instructions necessary
for constructing the solution.

For illustration, let's go back to the running example in Figure~\ref{ex:deprecated-method-other} (with the corresponding Javadoc comment in Figure~\ref{ex:code-hints}).
As mentioned in Section~\ref{sec:intro}, when building the corresponding component library for this example,
%
we must find the necessary
components that consume existing objects and create a \lstinline[breaklines=true]{Calendar} object, so that we can call method \lstinline[breaklines=true]{get} as suggested by the code hint.
Besides adding components that allow us to generate the required objects, we also need components for setting their
fields. In our example, we must call
\lstinline[breaklines=true]{calendar.setTime(date)} to set the calendar's date
based on the existing \lstinline[breaklines=true]{date} object.

Adding too many components to the library will
make the code generation task infeasible. In particular, we should
be able to differentiate between components that we can use
(we have or we are able to generate
all the necessary arguments and the current object for calling them), and those
that we can't because we can't obtain some of the arguments
and/or the current object. Adding the latter components to the library will
significantly slow down the code generation process by
adding infeasible programs to the search space.
We address these challenges by dynamically building the component library for each refactoring
such that it only contains components specific to that particular use case.

Throughout the seeding process, we keep track of the following sets:
\consumableobjs (inputs to the code to be refactored, which need to be consumed by the refactoring),
\availabletypes (types for which we either have consumable objects or the corresponding generators to create them) and
\targettypes (types for which we must be able to generate objects). To start with, the \targettypes set
contains the types of the original code's outputs. 

For the running example, we start with:
  $\consumableobjs = \{\texttt{date}\},
  \availabletypes = \{\texttt{Date}\},
  \targettypes    = \{\texttt{int}\}$.
%
%
%
The objective of the seeding algorithm is to
add components to the library that make use of the \availabletypes to generate objects
of \targettypes. At the same time, we want to consume the objects from the
\consumableobjs set, i.e., the inputs of the original code.

The seeding algorithm for the CodeHints-library is provided in Figure~\ref{alg:seeding-core} and consists of three phases, which we discuss next.

\paragraph{{\bf Phase 1: Initialise with code hints}}
During the first phase, the initialisation, we add all the constants and instructions from the code hints to the library.
%
For our running example, the hints in Figure~\ref{ex:code-hints} 
instruct us to add method ``\lstinline[breaklines=true]{int get(int field)}'' from class
\lstinline[breaklines=true]{Calendar} and constant ``\lstinline[breaklines=true]{Calendar.HOUR_OF_DAY}'' to our library.
As a side note, finding the right constants is a well known
challenge for program synthesis~\cite{DBLP:conf/cav/AbateDKKP18}, and thus the subsequent
synthesis process will always attempt to use the constants provided in the code hints before generating new ones.

Intuitively, we need to make sure that the Javadoc suggestions
are realisable as captured by Definition~\ref{def:realizable}, where
$\requiredtypes(\allowbreak method)$ refers to the types of the objects required to call $method$ (i.e., the types corresponding to its arguments and current object).
%
In our running example, $\requiredtypes(\texttt{get}) = \{\texttt{int, Calendar}\}$
given that, in order to call \lstinline[breaklines=true]{get}, we must
provide an argument of type \lstinline[breaklines=true]{int} and a current
object of type \lstinline[breaklines=true]{Calendar}.
Method \lstinline[breaklines=true]{get} is not realisable as the library
doesn't contain any generator for \lstinline[breaklines=true]{Calendar}.
Consequently, \lstinline[breaklines=true]{Calendar} is added to
\targettypes, resulting in:
$\targettypes = \{\texttt{int, Calendar}\}$.

\begin{definition}[Realisable method]\label{def:realizable}
Method \var{i} is {\em realisable} iff $\forall t \in
\requiredtypes(i). t \in \availabletypes \vee t~is~a~\mathit{primitive}~\mathit{type}$.
\end{definition}


One challenge in this phase is interpreting the
\lstinline[breaklines=true]{@code} blocks inside \lstinline[breaklines=true]{@deprecated} sections, which
we attempt to parse as Java expressions. In particular, we parse each hint as if it were invoked in the context of
the method to refactor, such that imports or the implicit \lstinline[breaklines=true]{this} argument
are considered during parsing.

%
In order to address the challenge that code hints are not always expressed as well-formed Java,
we customised the
GitHub Java parser
to accept undeclared identifiers and type names
as arguments. For instance, the Javadoc hint for deprecating \lstinline[breaklines=true]{boolean inside(int X, int Y)}
in Figure~\ref{ex:javadoc-hint} suggests using
\lstinline[breaklines=true]{contains(int, int)}, which would normally cause a parsing error as the \lstinline[breaklines=true]{int} type
appears in the place of argument names. In our setting, we accept this hint as valid.
There are still situations where our parser is too strict and fails to accept
some of the code hints. Additionally, there are scenarios where, while the Javadoc
does contain a useful code hint, it is not tagged accordingly with
the \lstinline[breaklines=true]{@code} tag. 



\begin{figure}
\begin{lstlisting}[mathescape=true,showstringspaces=false]
/* @deprecated As of JDK version 1.1,
* replaced by {@code contains(int, int)}.
*/
@Deprecated
public boolean inside(int X, int Y) { ... }
\end{lstlisting}
\caption{Javadoc hint example.}
\label{ex:javadoc-hint}
\end{figure}

\paragraph{{\bf Phase 2: Add generators for \targettypes}}

By generators we refer to constructors and any other methods returning
objects of that particular type.  In the algorithm in
Figure~\ref{alg:seeding-core}, once we added a generator for a new type, we
must add that type to \availabletypes.  Additionally, if the new generator
consumes any objects from \consumableobjs, we must remove them from the set.

For our running example, we must seed our component library with generators
for \lstinline[breaklines=true]{Calendar}.
We first scan all the public constructors of class
\lstinline[breaklines=true]{Calendar} and all the public methods from class
\lstinline[breaklines=true]{Calendar} that return an object of type
\lstinline[breaklines=true]{Calendar}.  We find the following four options:
\begin{enumerate}
  \item \lstinline[breaklines=true]{static Calendar getInstance()} -- creates the object using the default time zone and locale.
  \item \lstinline[breaklines=true]{static Calendar getInstance(Locale)} -- creates the object using the default time zone and specified locale.
  \item \lstinline[breaklines=true]{static Calendar getInstance(TimeZone)} -- creates the object using the specified time zone and default locale.
  \item \lstinline[breaklines=true]{static Calendar getInstance(TimeZone, Locale)} -- creates the object with the specified time zone and locale.
\end{enumerate}

Out of the four methods, only the first one is realisable.  Conversely, in
order to call the second method, we would need to generate an object of type
\lstinline[breaklines=true]{Locale}, for which we don't have a corresponding
component in the library, and the same applies to the last two methods.
Thus, we only add the first method to the CodeHints-library.

\paragraph{{\bf Phase 3: Add transformers for \targettypes}}
The third and last phase adds  transformers for the target types.  By
transformers we refer to methods that modify the value of an instance
variable.  Given our objective to generate objects for the \targettypes
while consuming objects from \consumableobjs, we prioritise transformers
that consume such objects.  For instance, for our running example there are
16 public transformers for the \lstinline[breaklines=true]{Calendar} class.
However, instead of adding all of them to the CodeHints-library, we
prioritise those that consume the \lstinline[breaklines=true]{date} object.
There is only one such transformer \lstinline[breaklines=true]{void setTime(Date date)}.
Once we added any new transformers to the library, we
remove the consumed objects from \consumableobjs.

Notably, all the components that we add to the library must be accessible
from the current location.

\begin{figure}
\removelatexerror
\begin{algorithm}[H]
\SetAlgoLined
\KwOut{CodeHints-library}
// {\bf Phase 1: Initialise}\\
Add constants and instructions from the code hints to the library\;
\For{each unrealisable instruction $i$ in library}{
  $\targettypes = \targettypes \cup \requiredtypes(i)$\;
}
$\newline$
// {\bf Phase 2: Add generators for \targettypes}\\
\For{each $t \in \targettypes$ for which there is no generator}{
  Add realisable generators for $t$ to library\;
  $\availabletypes = \availabletypes \cup \{t\}$\;
  Remove consumed objs from \consumableobjs\;
}
$\newline$
// {\bf Phase 3: Add transformers for \targettypes}\\
\While{$\consumableobjs\neq\emptyset$}{
Add realisable transformers to the library for types in
\targettypes that consume objects from \consumableobjs\;
Remove consumed objs from \consumableobjs\;
}
\end{algorithm}
 \caption{Seeding algorithm for the CodeHints-library}
\label{alg:seeding-core}
\end{figure}

It may be the case that there is no realisable generator for a target type,
or no transformers for the \targettypes that can consume all the
\consumableobjs.  When we finish exploring all the available
methods, we exit the corresponding loops in phases~2 and~3.  As a
consequence, the synthesiser may fail to generate a valid refactoring from
the CodeHints-library.
%
%
A possible future direction is allowing unrealisable generators and
transformers to be added to the library and iterating phases 2 and 3 a given
number of times (potentially until reaching a fixed point).
%

\subsection{Types-library} \label{sec:extend-library}

For cases when code hints are not provided, in addition to the
CodeHints-library, we also provide a component library that is populated
based only on the type signature of the deprecated method.  Next, we
describe how this Types-library is being built.

For the CodeHints-library, the Javadoc code hints guidance results in the code hints being
used to collect the \targettypes set during the initialisation phase.
If we don't have access to code hints and we are seeding solely
based on types, we start with the \targettypes set containing
the types 
of the outputs of the original code. 
Phases 2 and 3 of the seeding algorithm are the same as those in Figure~\ref{alg:seeding-core},
where we attempt to add generators and transformers for the \targettypes to the library.


%
Compared to the CodeHints-library, the seeding process for the Types-library
may end up missing critical types provided by the Javadoc code hints.  For
instance, in our running example, the \lstinline[breaklines=true]{Calendar}
class is only mentioned by the code hints and would not be included in the
seeding of the Types-library.  One possibility for adding
\lstinline[breaklines=true]{Calendar} to the target types without considering
the Javadoc hints, would be to add all the classes from the
\lstinline[breaklines=true]{java.util} package, which contains
\lstinline[breaklines=true]{Date}.  However, doing so would result in a very
large component library, most likely outside the capabilities of existing
synthesis techniques.

\section{Design and implementation choices} \label{sec:implementation}



\subsection{Abstract classes and interfaces}\label{sec:abstract}

The programs that need to be checked for equivalence may refer to
abstract classes and interfaces. These are by default instantiated
using the mocking framework Mockito.
Alternatively, we also curate a
list of explicit constructors
of subclasses to be used in favour of Mockito mocks for certain types, and users
have the option to extend this list with a custom configuration.
%
This is particularly useful when attempting to avoid constructors that are not amenable to fuzzing, such as
collection constructors that take an integer capacity argument, where an
unlucky fuzzed input might lead to an out of memory error.



%


\subsection{Observable state}\label{sec:observable}

Our equivalence check uses the Java Reflection API to observe the side effects
of programs, such as assigning new values to fields. As a consequence, we cannot
observe effects that are not visible through this API, such as I/O operations or
variables maintained in native code only. I/O operations could be recorded using
additional bytecode instrumentation in future work, but for the scope of our current
implementation, if programs only differ in such 
side effects, our engine is prone to producing a refactoring that it's not fully equivalent to the original.

\subsection{Instrumentation and isolation}

We use reflection to invoke the original method to refactor with fuzzed inputs,
as well as our synthesised refactoring candidates. This allows us to dynamically
invoke new candidates without the need for compilation.
Regarding static fields, Java
allows to load the same class in different class loaders, which creates
independent copies of its static fields.

In order to fully isolate the program state during the execution of the original and refactored code from each other, we load all involved classes in separate
class loaders. These class loaders are disposed immediately after the current set
of fuzzed inputs are executed, and new class loaders are created for the next
inputs. 
Classes that do not maintain a static state are loaded in a shared
parent class loader to improve performance.

The OpenJDK Java virtual machine implementation does not allow us to load
classes contained in the \lstinline[breaklines=true]{java.lang} package in such an isolated class
loader. For such classes we cannot apply refactorings that depend on static
fields, as they cannot be reset and will retain the state of previous
executions. Instead of observing the effect of one isolated refactoring
candidate, we would observe their accumulated effect, which would be incorrect.
For the scope of our
experiments this did not pose a problem, since most of the affected classes in
the \lstinline[breaklines=true]{java.lang} package do not maintain a static state, and the ones who do
were irrelevant to our refactorings.

\subsection{Checking aliasing}

While in Section~\ref{sec:equiv}, we discussed aliasing preservation in terms of
preserving the equivalence classes induced by the aliasing relation,
in our implementation we enforce a simpler but stricter  than necessary check.
In particular, all the objects
that are being referenced during the code's execution
are assigned increasing symbolic identifiers.
Then, we enforce aliasing preservation (condition (4) in Definition~\ref{def:prog-equiv-add}),
by checking that, for all variables defined by both the original and the refactored code,
the objects they reference have the same symbolic id.

For illustration, in the original code in Example~\ref{ex:aliasing}, the object of type \lstinline[breaklines=true]{Dimension} referenced by variable \var{dim1}
is assigned symbolic value \var{s_1}, whereas the object referenced by \var{dim2} is assigned symbolic value \var{s_2}.
In a similar manner, the object referenced by both \var{dim1} and \var{dim2} in the refactored code is assigned value \var{s_1}.
Then, the aliasing check fails as, in the original code, the object referenced by \var{dim2} has symbolic value \var{s_2},
whereas, in the refactored code, the object referenced by \var{dim2} has symbolic value \var{s_1}.
While this is a  stronger than needed requirement, it was sufficient for our experiments.
In particular, we didn't find any benchmark where a sound refactoring
was rejected due to it.



\section{Experimental evaluation}\label{sec:experimental-results}

We implemented the refactoring generation techniques in two tools called \tool and \llm, corresponding to the symbolic and the neural approach, respectively (available together with all the experiments at \url{https://github.com/pkesseli/refactoring-synthesis/tree/hanliang/dev}).

\subsection{Experimental setup}

Our benchmark suite contains 236 out of 392 deprecated methods in the Oracle
JDK 15 Deprecated API documentation~\cite{OracleJdk15DeprecatedAPI}.
We included all the deprecated methods except those that were deprecated without a replacement (e.g. \lstinline[breaklines=true]{java.rmi.registry.RegistryHandler.registryImpl(int)}),
methods inaccessible by non-JCL classes (i.e., all the finalize methods, which are called by the garbage collector, not user code, meaning that we can't modify their calls),
methods that only perform I/O (as our equivalence check doesn't include I/O side effects),
and native methods.

For \llm, we use Claude 2.1 and Claude 3, denoted as \llma and \llmb, respectively.
Claude 2.1 and Claude 3 are proprietary LLMs likely to be very
large (1T+ parameters), which  have been shown to be among the highest performing on coding tasks~\cite{DBLP:journals/corr/abs-2405-11514,jiang2024surveylargelanguagemodels,hou2024comparinglargelanguagemodels,murr2023testingllmscodegeneration}.
We accessed the LLMs via Amazon Bedrock.
To make our results more deterministic, we use a lower temperature (i.e. less random) of $0.2$. We also repeated the experiments three times, and the results are summarized as averages.
For all engines, we bound the search by at most 500 inputs and 5 minutes per verification phase, and at
most 2 minutes per synthesis phase.

Experiments were performed on an Ubuntu
22.04 x64 operating system running in a laptop with 16\,GB RAM and
11th Gen Intel Core i7-11850H at 2.50\,GHz.  The JVM used was Oracle JDK 15.02.


For \tool, if code hints are present, we start with the CodeHints-library and fall back to the Types-library if the synthesis phase (as described in Section~\ref{sec:synthesis}) times out for the CodeHints-library; if no code hints exist then we use the Types-library.

\subsection{Results}


\begin{table}[h]

\resizebox{\columnwidth}{!}{
\begin{tabular} {|l|r|r|r|r|r|}
\hline
Configuration & \checkmark & \xmark & \lightning & \% & $\diameter$ runtime (s) \\ \hline
\tool       &        7 &    82 &         16 & 6 & 213.7 \\
\llma                     &        12 &     91 &         2  & 11 & 197.14 \\
\llmb                     &        10 &     93 &         2  & 9 & 203.32 \\
Best Virtual Engine    &        15  &         77 &      13 &  14 & 195.9 \\
\hline\hline
\tool  (CH)     &        104 &     19 &   8         & 79 & 220.8 \\
\llma  (CH)  &             93.3 &     36.7  &  1         & 71 & 212.19 \\
\llmb   (CH) &             94 &     36  &  1         & 72 & 217.9 \\
Best Virtual Engine (CH) &           107 &     13  &  11         & 82 & 210.28 \\
\hline\hline
\end{tabular}
}
\caption{Experimental results for all benchmarks.}
\label{tab:configuration-results}
\end{table}

Table~\ref{tab:configuration-results} provides an overview of the number of
sound refactorings (\checkmark), missed refactorings (\xmark), unsound
refactorings (\lightning), the percentage of sound refactorings (\%)
produced per configuration, as well as the average runtime per refactoring.
For the symbolic engine, refactorings are guaranteed to compile so missed refactorings are those that failed our equivalence check, whereas for the neural engine,
they either didn't compile or failed the equivalence check.
Unsound refactorings are
those that passed the equivalence check, but were manually detected by us as not being equivalent to the original code.
We describe the reasons why this can happen later in this section.
The average runtime includes both the time to generate a refactoring and to verify it.

We split our dataset into benchmarks where code hints could be extracted from the Javadoc, and benchmarks without code hints.
The first four rows provide the results for benchmarks without code hints, whereas the last four (marked with ``(CH)'') show the results for those
benchmarks where code hints were present. The rows denoted by ``Best Virtual Engine'' count all benchmarks (with and without code hints, respectively)
solved by at least one engine.

The results support our hypothesis that code hints are very valuable for automating the refactoring process.
For all engines, benchmarks with code hints have a considerably higher success rate than those without code hints, with the best virtual
engine solving 82\% of the benchmarks with code hints. In the absence of code hints, all the engines are struggling, solving only a few benchmarks.

To understand the discrepancy between the number of missed/unsound refactorings with the without code hints,
we next investigate the main reasons for such refactorings.


%
%

\paragraph{Missed refactorings}
%
%
%
For the symbolic engine, the majority of the missed refactorings were due to fuzzing timeouts, which are likely
caused by the component libraries missing some
instructions needed in the synthesis phase.
As explained in Section~\ref{sec:components-seeding}, one option
for increasing the size of our libraries
is allowing unrealisable generators and transformers to
be added.

For the neural approach, we could only observe that the LLMs were unable to generate the refactoring from the provided prompt.

For both engines, when refactoring methods that eventually run native code, we encountered crashes of the verifier, which, in order to be conservative, we counted as overall missed refactorings.
The reason for this is the fact that we use reflection to produce counterexamples,
and some may violate internal invariants.
If they execute methods that eventually call native code, this can lead to the entire JVM
crashing rather than e.g. throwing an exception. 

\paragraph{Unsound refactorings}
%
%
In several cases, refactorings that are not equivalent to the original code managed to pass our verifier.
While we expected to run into this problem because of the nature of fuzzing (the fuzzer may miss counterexamples that distinguish the behaviour of the original from the refactored code),
we also encountered other problems:


Unobservable state (discussed in Section~\ref{sec:observable}): I/O operations, static state in the boot class loader and
native methods are not observable by our equivalence predicate implementation,
since we rely on reflection to examine objects inside the Java runtime. As
a consequence, we cannot distinguish programs that only differ in these aspects.

Abstract methods:
There are classes in the JCL without any
existing implementation (e.g. \lstinline{javax.swing.InputVerifier}), and thus no concrete
method against which to verify the equivalence between the original and the refactored code. For those, the symbolic engine will
generate a no-op. We've been very conservative here, and counted this scenario as ``unsound'' because
it doesn't match human intent for the deprecated method.

Insufficient counterexamples: Our fuzzing-based equivalence check is inherently
incomplete, and for some benchmarks we do not explore sufficient counterexamples
to identify unsound candidates. An example
is \lstinline[breaklines=true]{javax.swing.JViewport#isBackingStoreEnabled}, where only a single input
out of the $2^{32}-1$ possible input values will trigger an alternate code path.
As a second exemplar, method \lstinline[breaklines=true]{java.rmi.server.RMIClassLoader#loadClass}
accepts a string as an input, but will throw when given any string
that is not a valid class name. 
It is very unlikely that our fuzzer will randomly produce a string that matches a valid class
name.

\paragraph{Discussion}

The symbolic engine benefits significantly from code hints, as hints seed
the component library, making it less likely that needed instructions are
missing.  For the neural engine, we hypothesise that, by providing
additional context to the LLM, code hints are aiding the code generation
task.

Benchmarks with code hints also have fewer unsound results.  Our hypothesis
is that both engines are much more likely to generate the expected
refactoring before generating other candidate refactorings that may trick
our equivalence checker.

\subsection{Research questions}

{\bf (RQ1) Can the refactoring of deprecated Java APIs be automated?}
Yes, the refactoring of deprecated Java APIs can be automated if code hints are added to the JDK,
evidenced by the 82\% success rate for those benchmarks.
This automation could greatly assist engineers with the code migration task.

\noindent
{\bf (RQ2) Do code hints help the generation of refactorings?}
The performance of \tool, \llma and \llmb improves considerably when code
hints are present.  All of them barely manage to solve any benchmark without
code hints.  When code hints are present, all engines solve at least 71\% of
benchmarks, with the best virtual engine solving 82\%.  For the symbolic
engine, they enable the effective seeding of the instruction library,
whereas for the neural engine, they provide additional context to the LLM.
In summary, without code hints, Java's complexity makes the refactoring
of deprecated Java APIs very hard for both for the
symbolic and neural engines.

\noindent
{\bf (RQ3) How do the symbolic and the neural approach compare against each other?}
The symbolic and the neural engines have very similar performance (where the symbolic engine has a smaller computational cost).



When code hints are present, as shown in
Table~\ref{tab:configuration-results}, the symbolic approach does slightly
better than both \llma and \llmb.  This supports the intuition that, if
there is enough information about the solution to effectively prune the
solution space (in our case, when code hints are present),
%
then the symbolic approach works well. 
One example of a benchmark that was solved by \tool but where both \llma and
\llmb failed to find a solution is the running example in
Figure~\ref{ex:deprecated-method-other}.  In all our runs, the LLMs failed
to generate \lstinline[breaklines=true]{calendar.setTime(date)}.

When code hints are not present, the neural engine does mar\-gin\-al\-ly better
than the symbolic approach.
%
%
%
%
Compared to the symbolic engine, the neural one is able to provide correct
refactorings for deprecated concurrency primitives (e.g.,
\lstinline{weakCompareAndSet} in
\lstinline[breaklines=true]{java.util.concurrent.atomic.AtomicReference}).
These methods call native methods, for which we cannot observe side-effects
(Section~\ref{sec:observable}).  Consequently, these are not captured by
the counterexamples returned by the fuzzer, and are thus not taken into
consideration by the symbolic engine during the synthesis phase.  For the
neural approach, the counterexamples are less critical, and the LLM can
obviously generate the correct code even though they are incomplete.

The neural engine is also able to handle benchmarks that require special
constants, such as
\lstinline[breaklines=true]{java.net.URLDecoder.decode(s)} described in the
introduction.

\section{Threats to validity}

\paragraph{Selection of benchmarks}
All our benchmarks are Java methods deprecated in the Oracle JDK 15.
The JDK is extremely well known to Java developers, and a lot
of Java application code evolves similarly.
%
However, our claims may not extend to other programming languages.


\paragraph{Quality of refactorings} Refactorings need to result in code that
remains understandable and maintainable.  It is difficult to assess
objectively how well our technique does with respect to this subjective
goal.  This threatens our claim that refactoring of deprecated Java APIs
be automated.

%
We manually inspected the refactorings obtained with
both engines and found them to represent sensible transformations.

\paragraph{Efficiency and scalability of the program synthesiser}
We apply program synthesis and fuzzing.  This implies that our broader claim
is threatened by scalability limits of these techniques.  While for the
majority of our experiments the synthesiser was able to find a solution,
there were a few cases where it timed out, either because it could not
generate a candidate, or it could not verify it.  Component libraries with a
diverse range of component sizes may help mitigate this effect.

\paragraph{Prompt engineering for Claude}
In our current experiments, we engineered prompts for the Claude LLMs.
While we made best efforts to follow the official prompt engineering
guidelines\footnote{https://docs.anthropic.com/en/docs/prompt-engineering},
which presents prompt design techniques to improve model performance, there
might be better ways of composing it.  This might invalidate our claim that
the symbolic and neural techniques deliver roughly the same performance.

\section{Related Works}

\paragraph{Program refactoring}

We first discuss works on the refactoring of deprecated instances.  The work
of Perkins directly replaces calls to deprecated methods by their
bodies~\cite{DBLP:conf/paste/Perkins05}, but we argue this conflicts with the intent of language designers.
Moreover, it can introduce concurrency bugs as inlining calls to
deprecated methods can cause undesirable effects if the original function
was synchronised.  While it is not possible for two invocations on the same
object of the synchronised original method to interleave, this is not
guaranteed after inlining the method's body.  Notably, the authors of~\cite{10.1007/978-3-642-14107-2_11} show how refactorings in concurrent programs can inadvertently introduce concurrency issues by enabling new interactions between parallel threads.


A related class of techniques aim to adapt APIs after library updates.  Such
techniques automatically identify change rules linking different library
releases~\cite{DBLP:conf/icse/WuGAK10,DBLP:conf/kbse/Huang0PW021}.
Conversely to our work, a change rule describes a match between methods
existing in the old release, but which have been removed or deprecated in
the new one, and replacement methods in the new release.  However, they do
not provide the actual refactored code.
In~\cite{DBLP:journals/tse/LeeWCK21}, Lee et al.~address the problem of
outdated APIs in documentation references.  Their insight is that API
updates in documentation can be derived from API implementation changes
between code revisions.  Conversely, we are looking at code changes, rather
than documentation.  Other works focus on automatically updating API usages
for Android apps based on examples of how other developers evolved their
apps for the same
changes~\cite{DBLP:conf/issta/FazziniXO19,DBLP:conf/iwpc/HaryonoTKSML0J20}.
Furthermore, \cite{DBLP:journals/ese/HaryonoTLJLKSM22} improves
on~\cite{DBLP:conf/iwpc/HaryonoTKSML0J20} by using a data-flow analysis to
resolve the values used as API arguments and variable name denormalization
to improve the readability of the updated code.  As opposed to these works,
which rely on examples of similar fixes, our approach uses Javadoc code
hints.





Several rule-based source-to-source transformation systems exist that allow users to define transformation rules based on the program's syntax~\cite{stratego,txl}. Unlike our approach, these systems require users to manually specify the transformation rules. In the context of code translation, a few works have explored the automatic generation of such rules using static analysis, but these efforts have been specialised to C-to-Rust translation~\cite{DBLP:conf/cav/ZhangDYW23,10.1145/3485498}.


Search-based approaches to automating the task of software refactoring,
based on the concept of treating object-oriented design as a combinatorial
optimisation problem, have also been proposed~\cite{search1,search2}.  They
usually make use of techniques such as simulated annealing, genetic
algorithms and multiple ascent hill-climbing.
%
%
While our technique is search based, the search is guided by types, code
hints, as well as counterexamples, discovered by testing.

Closer to our work, several refactoring techniques make explicit use of some
form of semantic information.
%
%
Khatchadourian et al.~use a type inference algorithm to automatically
transform legacy Java code (pre Java 1.5) to use the \lstinline[breaklines=true]{enum}
construct~\cite{sawin}.
%
%
Other automated refactoring techniques aim to transform programs to use a
particular design pattern~\cite{chris,bae}.
%
%
%
%
%
Steimann et al.~present Constraint-Based
Refactoring~\cite{Steimann2011,Steimann2012Pilgrim,Steimann2011KollePilgrim},
where given well-formedness logical rules about the program are translated
into constraints that are then solved to assist the refactoring.
%
%
Fuhrer et al.~implement a type constraint system to introduce missing type
parameters in uses of generic classes~\cite{DBLP:conf/ecoop/FuhrerTKDK05}
and to introduce generic type parameters into classes that do not provide a
generic interfaces despite being used in multiple type
contexts~\cite{DBLP:conf/icse/KiezunETF07}.
%
%
%
%
%
%
Kataoka et al.~use program invariants (found by the dynamic tool Daikon) to
infer
%
%
whether specific refactorings are
applicable~\cite{Kataoka:2001:ASP:846228.848644}.  Finding invariants is
notoriously difficult.  Moreover, the technique is limited to a small number
of refactorings and does not include the elimination of deprecated
instances.  In~\cite{conf/sigsoft/GyoriFDL13}, Gyori et al.~present the tool
{\sc LambdaFicator}, which automates two pattern-based refactorings.  The
first refactoring converts anonymous inner classes to lambda expressions.
The second refactoring converts {\tt for} loops that iterate over
Collections to functional operations that use lambda expressions.  The {\sc
LambdaFicator} tool~\cite{DBLP:conf/icse/FranklinGLD04} is available as a
NetBeans branch.

Our techniques does not expect any
precomputed information such as logical well-formedness properties or
invariants. Instead, we make use of information that is already
available in the original program and Javadoc, namely code hints and
type information. Moreover, we explore the space of all potential
candidate programs by combining techniques from type-directed,
component-based and counterexample-guided inductive synthesis.

\paragraph{Symbolic program synthesis}

CEGIS-based approaches to program synthesis have been previously used for
program transformations such as superoptimisation and
deobfuscation~\cite{DBLP:conf/icse/JhaGST10}.
In~\cite{DBLP:journals/corr/abs-1712-07388}, David et al.~present an
automatic refactoring tool that transforms Java with external iteration over
collections into code that uses Streams.  Their approach makes use of formal
verification to check the correctness of a refactoring.  Cheung et
al.~describe a system that
transforms fragments of application
logic into SQL queries~\cite{DBLP:conf/pldi/CheungSM13} by using a
CEGIS-based synthesiser to generate invariants and postconditions
validating their transformations (a~similar approach is presented
in~\cite{DBLP:conf/cc/IuCZ10}).  While our approach is also CEGIS-based,
we are guided by code hints and types to efficiently prune the
search space.  

%
Type information has been extensively used in program synthesis to guide the
search for a
solution~\cite{DBLP:conf/sfp/Katayama05,DBLP:conf/pldi/FeserCD15,DBLP:conf/pldi/OseraZ15,DBLP:journals/pacmpl/LubinCOC20,DBLP:journals/pacmpl/YamaguchiMDW21}.
In our work, we combine type information with code hints.
%
%
%
Another direction that inspired us is that of component-based
synthesis~\cite{DBLP:conf/icse/JhaGST10,DBLP:conf/pldi/GulwaniJTV11,DBLP:conf/popl/FengM0DR17},
where the target program is generated by composing components from a
library.  Similarly to these approaches, we use a library of components for
our program generation approach.  However, our technique uses information
about types and code hints to build the component library, which is specific
to each refactoring.


%
%

\paragraph{Large Language Models}
LLMs have been used successfully for code generation tasks,
with applications ranging from code
completions~\cite{llmsforcodecompletion,ni2023lever,codegenclasslevel,Ding2024cocomic},
translations~\cite{tang-etal-2023-explain,lostintranslation} to
repository-level generation~\cite{zhang2023repocoder} and general software
engineering tasks~\cite{yang2024sweagent}.  Those models are typically
pre-trained on vast amounts of data, fine-tuned for specific tasks, and
require advanced prompt engineering~\cite{jiang2024survey}.
Among the well-known LLMs, GPT-4~\cite{openai2024gpt4}, Claude2.1~\cite{claude},
Claude3~\cite{claude}, LLaMa~\cite{touvron2023llama} are general-purpose
models covering a diverse set of language-related applications; there are
also models pre-trained/fine-tuned specifically for code generation and
programming tasks, such as CodeLLaMa2~\cite{roziere2024code},
StarCoder2~\cite{lozhkov2024starcoder},
DeepSeek-Coder~\cite{guo2024deepseekcoder}, and GrammarT5~\cite{grammart5}.
%
In this work, we have applied the Claude family of LLMs to our refactoring
task, with the goal of investigating how much it benefits from Javadoc code
hints.  





\section{Conclusions}

In this paper, we investigated the benefits of Javadoc code hints when
refactoring deprecated methods.  For this purpose, we designed and
implemented a symbolic and a neural automatic program refactoring techniques
that eliminate uses of deprecated methods.  In our experiments, both engines
demonstrate strong performance when code hints were present, and fare much
worse otherwise, leading us to conclude that code hints can boost the
automation of refactoring code that uses deprecated Java APIs.

\paragraph{Acknowledgements}
Cristina David was supported by the Royal Society University Research Fellowship URF{\textbackslash}R{\textbackslash}221031.

\bibliographystyle{ACM-Reference-Format}
\bibliography{document}

\end{document}